\documentclass[final,5p,times,twocolumn]{elsarticle}

\usepackage{amssymb}
\usepackage{amsmath}

\usepackage{lineno}

\usepackage{lipsum}

\usepackage[dvipsnames]{xcolor}

\journal{Nucl. Instrum. Methods Phys. Res., Sect. A}

\usepackage{threeparttable}
\usepackage{tabularx}
\usepackage{booktabs}
\usepackage{hyperref}
\usepackage{listings}
\usepackage{siunitx}
\usepackage[version=4]{mhchem}

\newcommand{\iso}[2]{\ce{^{#1}#2}}
\newcommand{\goodness}{\ensuremath{\delta^{2}\,}}

\begin{document}

\begin{frontmatter}



\title{Track reconstruction with ACTAR TPC: a transfer-reaction use case}

\author[1]{M. Lozano-González} 
\author[1]{J. Lois-Fuentes} 
\author[2]{T. Roger} 
\author[1]{I. Blanco-Calviño}
\author[1]{B. Fernández-Domínguez}

\affiliation[1]{organization={IGFAE and Dept. Particle Physics, University of Santiago de Compostela},
            addressline={r/Xoaquín Díaz de Rábago}, 
            city={Santiago de Compostela},
            postcode={E-15706}, 
            state={A Coruña},
            country={Spain}}
\affiliation[2]{organization={GANIL, CEA/DRF-CNRS/IN2P3},
            addressline={Bd. Henri Becquerel, BP 55027}, 
            city={Caen},
            postcode={F-14076}, 
            country={France}}

\begin{abstract}
A unified framework for track reconstruction in the ACTAR Time Projection Chamber (ACTAR TPC) is presented, designed to address the specific challenges of direct-reaction and resonance-elastic scattering experiments involving exotic nuclei. The \textsc{actroot} software package enables efficient semi-online analysis by integrating a continuity-based clustering algorithm followed by a modular filtering procedure to reduce noise, delta-electron contamination, and track fragmentation. The software was benchmarked with experimental data taken with a secondary \iso{20}{O} beam delivered by the LISE3 spectrometer at GANIL. The algorithm achieves reconstruction efficiencies exceeding 90\% across various event topologies. The system provides accurate particle identification and reaction-point determination, facilitating the extraction of key physical observables. These results demonstrate the framework’s suitability for its application to direct reaction experiments in cubic active-target TPCs without magnetic fields.
\end{abstract}




\end{frontmatter}



\section{Introduction and Physics motivation}
\label{sec1}

Understanding nuclei with extreme proton-to-neutron ratios is a key objective in modern nuclear physics. Recent experimental advancements have granted access to isotopes that were previously out of reach, revealing novel behaviours that challenge existing nuclear models and drive theoretical progress. To refine these theories, precise experimental data is essential. Direct reactions have long served as a foundation for nuclear structure studies, but employing radioactive beams introduces obstacles such as low event rates and complex kinematics that can reduce experimental resolution.

The last twenty years have witnessed a rapid expansion of active-target time projection chambers aimed at overcoming these challenges, with remarkable examples being MAYA \cite{demonchy2007}, CENBG TPC \cite{blank2010}, AT-TPC \cite{bradt2017} and ACTAR TPC \cite{ACTAR_demostrator_ThomasRoger, ACTARTPC_commissioning}, driven primarily by improvements in detector technology, such as Micromegas \cite{giomataris1996} and thick-GEM \cite{shalem2006,shalem2006a} gas amplification devices. The low intensity of radioactive beams limits luminosity when using solid targets, and increasing target thickness degrades resolution, forcing a trade-off between data quality and experimental feasibility.

Active targets address this issue by simultaneously acting as both the target material and detection medium. This configuration allows precise determination of reaction vertices, ensuring that target thickness minimally interferes with energy measurements. Additionally, low-energy recoil particles can be directly observed without needing to pass through an inert target layer, preserving both energy and angular resolution. Furthermore, the extensive solid-angle coverage provided by active-target TPCs enhances luminosity, making them invaluable for studying rare isotopes. A further advantage lies in their versatility to operate in different geometries: cylindrical TPCs embedded in a magnetic field, as in the AT-TPC, or cubic TPCs without external fields, as in ACTAR TPC. These technological strides will continue to shape nuclear structure research, yielding the crucial data necessary to advance our understanding of exotic nuclei.

This article presents the analysis framework developed for the ACtive TARget and Time Projection Chamber (ACTAR TPC) \cite{ACTAR_demostrator_ThomasRoger, ACTARTPC_commissioning} detector, with particular focus on the track-reconstruction algorithms used in direct-reaction studies. During the first transfer experiments performed with ACTAR TPC, there was a strong need for an analysis framework capable of delivering fast and reliable online results to monitor the development of the experiments. As more experimental data became available, the necessity for a unified and solid analysis tool within the collaboration became evident. To tackle this need, \textsc{actroot} was developed \cite{zenodo_actroot}, and this article provides an extensive review of it. Section 2 provides an introduction to the detector and to the data gathered in transfer experiments. Then, Sections 3, 4 and 5 focus on the technical implementation of the different algorithms. Finally, Section 6 concludes by showing an application example with a real dataset.

\section{Detector and data description}\label{sec:detdatadescription}
ACTAR TPC is a cutting-edge detector developed through a collaboration between European laboratories and constructed at GANIL. It features a large active region (\qtyproduct{295 x 255 x 255}{\mm}) where charged particle trajectories are recorded through the ionisation of the gas. Electrons then drift towards a highly-segmented pad plane, consisting of \qtyproduct{128 x 128}{} squares with \qty{2}{\mm} side each (\textit{pads}), under the application of a constant electric field. A MICROMEGAS amplification system placed above the pad plane permits an electron avalanche, enhancing the signal from the ionisation electrons. 

\begin{figure}[!htb]
        \centering
        \includegraphics[width=\linewidth]{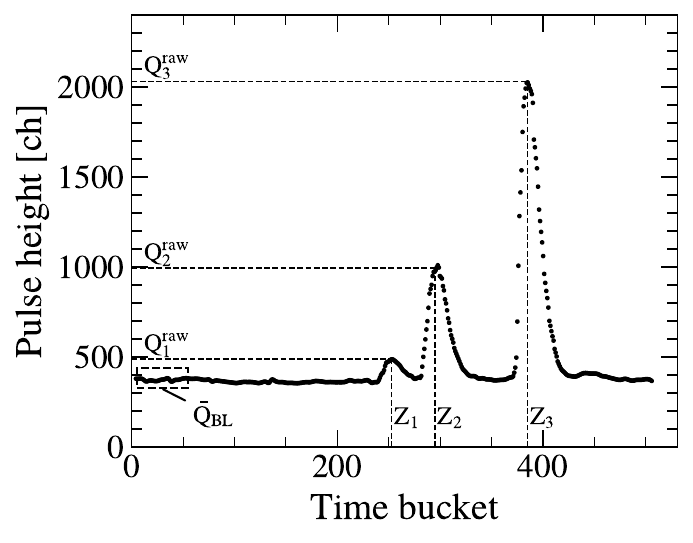}
        \caption{Example of signal extraction for a single pad. The baseline, estimated from the first 50 time buckets, is subtracted to obtain the corrected signal. Multiple peaks within the event are identified, each assigned a time bucket ($Z$) and a corrected charge value ($Q$).}
        \label{fig:pad_signal}
\end{figure}

The \textit{General Electronics for TPCs} (GET) device \cite{GET_POLLACCO201881} enables signal digitisation by means of a hierarchical acquisition structure. Every 64 pads are read out by an AGET unit, which performs charge integration, shaping, and discrimination. The signals are then stored in a switched capacitance array (SCA) comprising 512 time cells (\textit{time buckets}) with a tunable $f_{\text{write}}$ frequency. An AsAd board digitises the signals and groups together four AGET units. Four AsAds are further combined into a CoBo board, from which data is available to the user. Furthermore, four additional channels are added per AGET unit to perform subsequent noise suppression, with intricate analysis patterns already developed in \cite{Giovinazzo2016}. Figure \ref{fig:pad_signal} shows an example of the signal processing for a single pad. For each event, the average baseline is determined using the first 50 time buckets ($\overline{Q_{\mathrm{BL}}}$). This baseline is then subtracted from the raw signal to obtain the corrected pulse height, $Q_i = Q^{\text{raw}}_i - \overline{Q_{\mathrm{BL}}}$. As illustrated in the figure, multiple peaks can be present within a single electronic event. Each peak is characterised by its corrected amplitude and corresponding time bucket. In a later step, the time information is converted into a spatial coordinate along the drift direction ($Z$) using the known drift velocity.

After the pretreatment step, the input data for the present program consists of a vector of 4D points, whose coordinates represent a \textit{voxel}: $(X,Y,Z,Q)$. Additionally, signals from different auxiliary detectors, such as silicon arrays, scintillator crystals or beam-particle counters, are digitised and included in the data. These detectors can be crucial for obtaining information about particles escaping the pad-plane active area, as will be discussed later.

\begin{figure}[!htb]
    \centering
    \includegraphics[width=\linewidth]{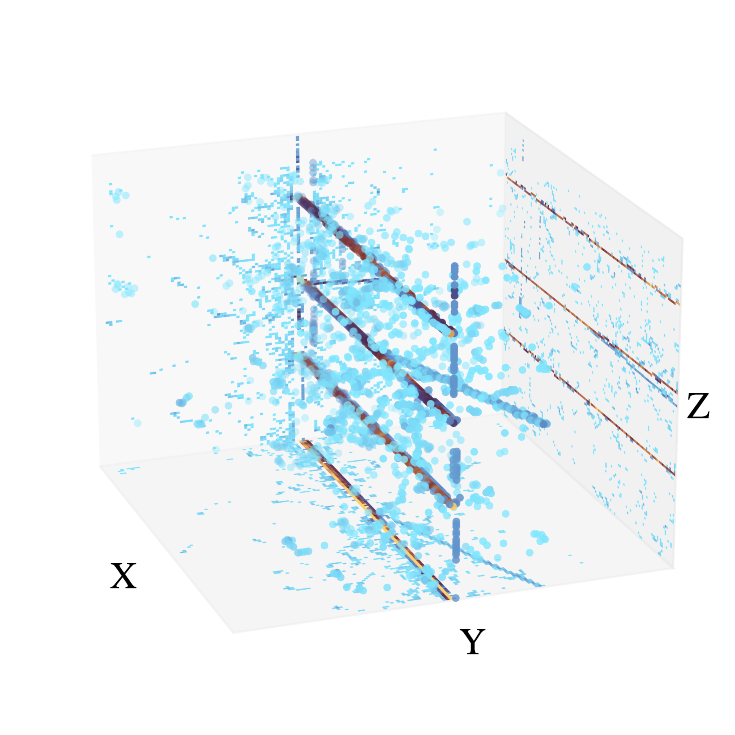}
    \caption{3D view of a voxel cloud for a typical event. Projections onto the different planes are also shown.}
    \label{fig:point_cloud}
\end{figure}

An example of the TPC data is provided in Figure \ref{fig:point_cloud}, where the voxel cloud is displayed. The 3D coordinate positions are represented by dots, while the colour scale indicates the voxel charge. Although graphically appealing, 3D representations are not well-suited for interpreting experimental data, as it is hard to visually infer particle trajectories. Instead, 2D projections are preferred, with the $XY$ and $XZ$ planes commonly employed, as also drawn in Fig. \ref{fig:point_cloud}. The $XY$ view allows clear identification of ejectile tracks, whereas the $XZ$ projection reveals time features, such as the \textit{pile-up} events, which occur when different beam particles enter during the same \textit{acquisition window}. This corresponds to the time window covered by the full reading of the SCA circular memory.

Within the scope of transfer reactions, in 2022 the first transfer experiment with ACTAR TPC was carried out. At the LISE spectrometer of GANIL (Caen, France), a \iso{20}{O} beam was delivered at approximately $35$A \unit{\MeV}. The drift volume was filled with a \qty{90}{\percent} \ce{D_2} and \qty{10}{\percent} \ce{iC_4H_10} mixture at \qty{950}{\milli bar}, aiming to measure $\iso{20}{O}(\iso{}{p}, \iso{}{d})$, $\iso{20}{O}(\iso{}{d}, \iso{}{t})$ and $\iso{20}{O}(\iso{}{d}, \iso{3}{He})$ \cite{loisfuentes2026} neutron and proton pick-up reactions, alongside the corresponding $\iso{20}{O}(\iso{}{p}, \iso{}{p^{\prime}})$ and $\iso{20}{O}(\iso{}{d}, \iso{}{d^{\prime}})$ elastic and inelastic scatterings. To detect the light particles, two \qty{500}{\micro\m}-thick silicon layers were placed at forward angles, covering \qtyrange{10}{50}{\degree} in $\theta_{\text{lab}}$, allowing access to the region where pickup transfer cross-sections are higher and also performing the particle identification (PID). Regarding the elastic scattering, a third layer was sited on the left flange, at \qty{10}{\cm} from the pad plane, spanning \qtyrange{60}{120}{\degree} $\theta_{\text{lab}}$ angles. Section \ref{sec:validation} will present the results of the PID obtained after track reconstruction in both kinematic regimes.

\begin{figure*}[!htb]
    \centering
    \includegraphics[width=1\textwidth]{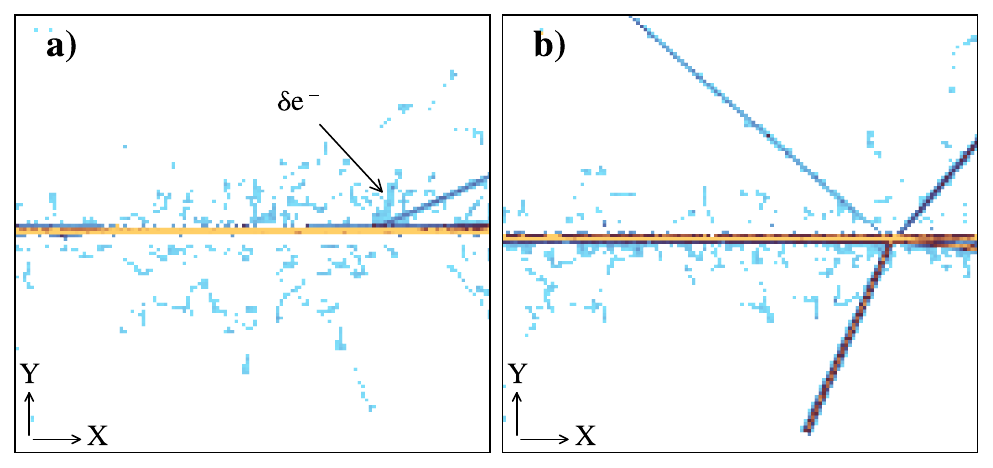}
    \caption{Pad-plane view of two challenging events: \textbf{a)} a binary event with a large $\delta e^-$ track, and \textbf{b)} a multifragmentation reaction.}
    \label{fig:example_events}
\end{figure*}

However, this experiment suffered from two major challenges owing to the particular experimental settings. First, to separate \iso{3}{He} and \iso{4}{He}, a high gain was set in the detector. A drawback of this is the significant presence of high-energy ionisation electrons (delta electrons, $\delta e^-$), which can travel considerable distances and be mistaken for reaction-product tracks. Figure \ref{fig:example_events}a) illustrates an example.

Another challenge arises from the high beam energy and the carbon component of the gas mixture. Under these experimental conditions, fragmentation reactions become the dominant trigger events, resulting in a broad spectrum of ejectile energies. Particularly problematic are high-energy tracks with such low energy deposition that they do not surpass the pad threshold, yielding fragmented tracks. As will be discussed in the next sections, reconstructing these tracks is especially arduous and, if a track is not recovered, it can lead to false multiplicities, misidentifying a multibody event as a binary reaction. A multifragmentation example is presented in Figure \ref{fig:example_events}b).

\section{Algorithm implementation}\label{sec:implementation}

The \textsc{actroot} program aims to be a polyvalent tool for analysing ACTAR TPC experiments, from converting the voxel data to extracting the physical observables. It is built in the versatile \lstinline|C++| programming language and uses the powerful \textsc{root} libraries \cite{root}. Two goals have driven its development: flexibility, enabling it to handle experiments carried out under different conditions, and ease of use, requiring minimal user interaction to run the code.

The design pattern was inspired by \textsc{nptool} \cite{Matta_2016}, another framework serving the nuclear physics community, but for which implementing the ACTAR TPC's intricate analysis was deemed too complex. Nonetheless, both frameworks leverage base classes to define the common structure of the analysis workflow (such as defining what a detector is, how a data structure must behave, or what an algorithm should do) and rely on \textit{manager} classes to operate between the different elements. This results in a highly modular code, facilitating user contributions, whether for bug detection or the addition of new components. Besides providing a basic analysis, \textsc{actroot} includes a set of useful physics and simulation classes to aid in the subsequent post-analysis stage: kinematic calculators, energy-loss table interpolators, and event inspectors, among others.

The program is organised into \textit{modes} that execute different tasks. The entry point is the conversion from the preprocessed data to our custom classes. This preprocessed data consists of a voxel vector with ancillary detector information appended, as discussed in Section \ref{sec:detdatadescription}. It is divided into three different types of spectra: \lstinline{TPCData}, representing the voxels from the TPC; \lstinline{SilData}, containing the silicon information; and \lstinline{ModularData}, which comprises any other accompanying data, such as trigger and scaler information.

While executing the conversion, two important operations are already performed. On the one hand, the \lstinline{TPCData} is clustered using the desired cluster algorithm, with Continuity \cite{maussThesis} or RANSAC \cite{fischler1981} as the two available options. A detailed description of the former is provided in Section \ref{sec:clustering} and \textsc{actroot} defaults to it. On the other hand, the \lstinline{SilData} is calibrated, and for the \lstinline{ModularData}, only a subset of variables is retained for further processing. This particular flow avoids having to re-execute the time-consuming cluster algorithm in case of a change in a calibration or modular variable selection.

Subsequently, a filter action is applied to the initial clusters. This is of paramount importance as, depending on the experiment, there might be high background conditions, or the preliminary tracks might be ill-formed (broken or mixed, for instance). This filter consists of different actions that are applied sequentially, modifying the shape of the preliminary clusters. Examples of necessary operations include merging broken clusters, removing noise, and finding a reaction point, which are described in Sec. \ref{sec:filtering}.

Lastly, a minimal physics output is created by merging back the \lstinline{TPCData} with the \lstinline{SilData} and \lstinline{ModularData}. Conditions on the silicon multiplicity, triggers, and track number are imposed to detect binary events. Employing their line representations, angles and track lengths are computed. This last part is highly experiment-dependent, and several parameters manage the different options, as Section \ref{sec:trackrec} will describe later on.  With this minimal output produced with \textsc{actroot}, the user can apply further custom analysis in separate standard \textsc{root} macros. 

\section{Cluster operations}

As described in the previous section, the two basic steps of the implemented algorithm are a clustering phase and a filter. The general design is an upgrade from the ClIMB algorithm \cite{JLF_Thesis}, which includes an alternative to the Continuity clustering (by using RANSAC), the use of a single volume for the tracking, a reorganisation and refinement of the filtering functions, and a centralised and systematic location to handle the critical operation parameters. 

\subsection{Continuity Clustering}\label{sec:clustering}
The clustering algorithm is the most computationally intensive step of the tracking process described in this paper. This is primarily due to the large voxel volume that defines our events, each consisting of $128 \times 128 \times 512 = 8.4 \times 10^6$ elements, although most are empty. The algorithm is designed for semi-online analysis during the experiment to provide immediate insight into its progress. Therefore, this step was carefully optimised to minimise processing time. The clustering method follows the principle of continuity, grouping adjacent voxels into the same cluster if they form a continuous track. Specifically, once a voxel is identified as part of a track, its neighbouring voxels are iteratively examined in all spatial directions (both adjacent and diagonal). This approach prioritises local regions with detected activity, reducing unnecessary searches and improving computational efficiency.

To reduce dimensionality, a rebin factor is usually applied to the initial vector of voxels in the Z coordinate. In this work, a rebinning factor of four was used, which therefore translates into an effective TPC size of $128\times 128\times 128$. The unbinned information is retained throughout the program and recovered in the track-reconstruction part (Sec. \ref{sec:trackrec}) to provide precise angle calculations.

With the initial vector of voxels $V_m$, the Continuity algorithm works in two complementary spaces: a vector of integers ($v_{m}$), which tracks whether the $m$-th voxel is assigned to a cluster (marked by a $-1$ sentinel value) or pending clustering (holding the voxel's position in $V$), and a 3D matrix ($M_{ijk}$) representing the spatial $(X,Y;Z)$ coordinates, whose non-null elements also store the voxel's index in $V$.

\begin{figure}[!htb]
    \centering
    \includegraphics[width=1\linewidth]{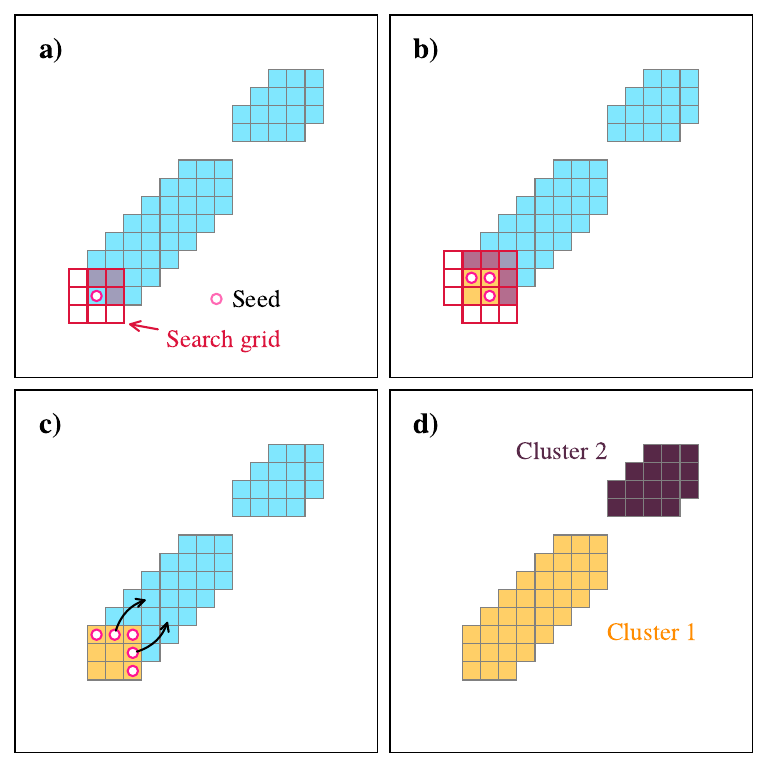}
    \caption{Two-dimensional view of the Continuity clustering algorithm acting on the voxel vector. See text for further details on the different steps.}
    \label{fig:continuity}
\end{figure}

The continuity algorithm consists of two fundamental operations: selecting a new seed and growing the cluster. These processes are illustrated in Figure \ref{fig:continuity} and explained in the following lines:
\begin{enumerate}
    \item Seed selection. The vector $v$ is searched for the first element with $v_m \neq -1$, whose value is retrieved in the $V$ vector and corresponds to an element in the $M$ matrix. This voxel constitutes the seed for the \textit{current} cluster and is shown by a white-filled circle in Fig. \ref{fig:continuity}a).

    \item First generation. In the $M$ matrix, all $3^3 -1$ adjacent directions around the seed are checked for voxel indexes. If these are found, a temporary vector storing their positions in $V$ is created, representing \textit{the first generation}. Their values in both $v$ and $M$ objects are set to $-1$ to prevent them from being processed again. In Fig. \ref{fig:continuity}a), this is illustrated in 2D by the red grid. Positions that fulfil the continuity condition are filled with a transparent red colour.

    \item Subsequent generations. Each voxel in the temporary vector is the seed for new generations, as shown in Fig. \ref{fig:continuity}b), and the process is recursively repeated until a generation returns an empty temporary vector. In the example of Fig. \ref{fig:continuity}, the cluster grows following the directions indicated by the arrows in the lower left panel, and stops when the continuity condition yields no new elements.

    \item Cluster validation. Once a cluster is identified, a user-defined threshold $N_{\text{min}}$ is applied to the number of voxels to determine whether the cluster is retained for further processing or considered noise. With typical thresholds in the \qtyrange{10}{20}{} range, this condition acts as a soft filter, primarily removing non-track structures, such as $\delta e^-$ hits. The resulting cluster is then fitted using an analytical 3D regression function, based on a modified version of the procedure described in Ref. \cite{jacquelin}, where the charge information from the ADC is used to weight the fit. This process defines a vector for each cluster and calculates the sum of the squared orthogonal distances of the voxels to the line (\goodness), which is used as a measurement of the fit's quality.

    \item Search for new clusters. The three steps above are iteratively repeated until all voxels in the original vector $V$ have been processed, or, equivalently, $v_m = -1\;\forall\;m$. In our example, the Continuity algorithm yields two separate clusters since the initial array had two non-continuous structures within it, as depicted Fig. \ref{fig:continuity}d).
\end{enumerate}

\subsection{Filtering}\label{sec:filtering}
\begin{figure*}[!htb]
    \centering
    \includegraphics[width=0.8\textwidth]{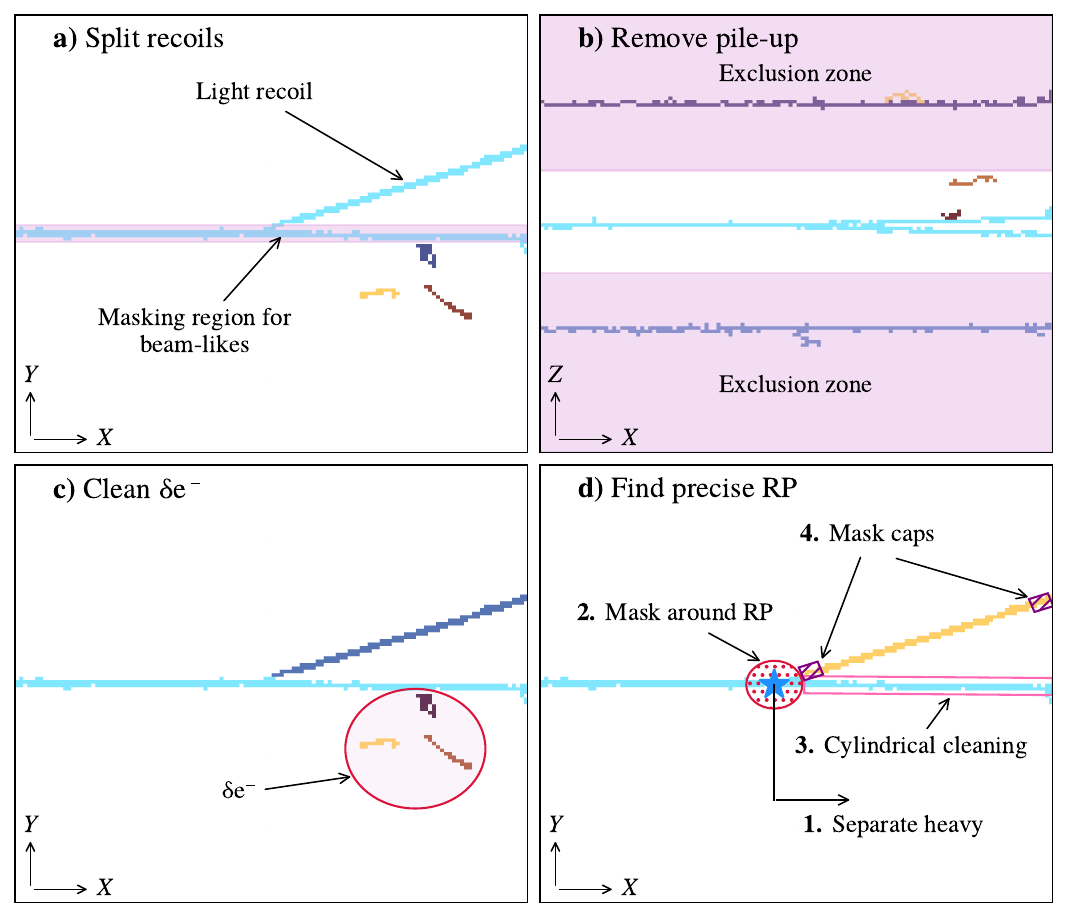}
    \caption{Sketch of the different actions involved in the filtering algorithm. Refer to Section \ref{sec:filtering} for further details.}
    \label{fig:filtering_all}
\end{figure*}

Once the clusters have been formed, they must undergo different transformations before being suitable for extracting the physical observables. These operations will depend on the experimental settings; in general, they involve cleaning the background, fixing broken tracks or splitting others that may be clustered together. The next subsections provide a detailed explanation of such actions, presented in their order of execution.

\subsubsection*{Splitting the recoils}
Considering that the continuity condition might be satisfied by multiple tracks, they could be recognised as a single one, yet with a high \goodness value. Therefore, this parameter is employed to decide whether the cluster should be split or not, given a threshold value. For clusters fulfilling this condition, their centre of gravity is computed, and a window is opened in the $YZ$ coordinates. Points outside this window are passed again to the clustering algorithm and, since their connecting points are now masked, they will form independent tracks, as illustrated in Fig. \ref{fig:filtering_all}a). This task usually involves beam clusters, with recoiling tracks appearing glued to the incoming particle. Depending on the recoiling particle angles, one or two new clusters can be formed.

\subsubsection*{Fixing broken clusters}
Conversely, the same physical track can be broken into different clusters if the particle has a low-energy deposit, which occurs for a wide range of kinematical scenarios or if there are some malfunctioning pads or electronic noise. This part of the algorithm runs an all-to-all cluster comparison and merges those that are compatible. To determine compatibility, several conditions must be fulfilled simultaneously: first, the gravity point-fit distance is evaluated for each element in a pair of clusters, and their maximum distance must be below a given threshold. Secondly, the opening angle between the two clusters must not exceed an upper limit (frequently below \qty{20}{\degree}), implying that both clusters are aligned. A further requirement is that their joint linear fit improves the initial fit goodness. 

\subsubsection*{Cleaning pile-up clusters}
In the standard direct reaction setup, the acquisition window typically lasts for \qty{40}{\micro\s}, which results in many beam particles being registered for the same event. Depending on the angle of the residuals and the proximity of beam-like clusters, it might be difficult to infer from which cluster ejectiles originate, complicating the reaction-point search in the next steps. Consequently, it is convenient to erase them at this stage. To this end, clusters fully contained within an exclusion area in the $XZ$ projection and nearly horizontal are deleted, as \autoref{fig:filtering_all}b) depicts. 
The beam clusters associated with the reaction particles are not removed, considering that the trigger delay is adjusted to centre them.

\subsubsection*{Cleaning $\delta$ electrons}
Lastly, one must erase the problematic $\delta$ electron clusters, which, depending on their energy, might be misidentified as ion tracks, especially when they are emitted at the end of the chamber and with a low angle. This operation erases all clusters whose \goodness surpasses a threshold, although a further constraint must be imposed since this could also delete broad tracks coming from heavy residues. By computing the cluster's $\sigma$ along its fit ($\sigma_{\parallel}$) and transversal ($\sigma_{\perp}$) dimensions, it can be unambiguously classified as an ion particle (which, being linear, would present $\sigma_\parallel \gg \sigma_\perp)$ or noise (with no preferred direction, $\sigma_\parallel \sim \sigma_\perp$). Only the latter are deleted. In \autoref{fig:filtering_all}c), the highlighted clusters will be erased after this operation. Lastly, a final gate on the number of voxels is imposed on all clusters to delete any remaining background, with an optimal threshold of \qty{21}{} voxels for this particular experiment.

\subsubsection*{Finding a preliminary reaction point}
The last goal of filtering involves finding a unique reaction point for the event, as this removes any residual uncorrelated tracks with the one that triggered the event. The first step consists of classifying the tracks into beam-like and recoil-like categories. Beam clusters are characterised by being parallel to the X axis (although a small opening angle is allowed, $< \qty{5}{\degree}$ in this experiment) and close to the beginning of the pad plane (distance $< 15$ pads). The remaining tracks are automatically tagged as recoils.

Once classified, all possible beam and recoil cluster pairs are intersected, and their preliminary reaction points (RP, or vertex) are computed. However, as two 3D lines in general do not intersect, the RP is taken as the average of the two closest approach points between both lines. Next, a set of conditions is applied to validate the different intersection points. First, the common perpendicular (the shortest line connecting two skew lines) is limited so that only correlated clusters are considered. Secondly, RPs are grouped together within a given radius to correctly treat $N$-body reactions, where there might be several RPs for the same beam track. Lastly, another classification occurs between tracks that have an RP and those that do not. The latter clusters are erased from the event, allowing us to proceed with a detailed analysis of the tracks coming from the direct reaction.

\subsubsection*{Fine analysis of tracks}
The filtering concludes by returning the tracks ready for the extraction of physics observables. To achieve this, a range of different corrections must be applied, as \autoref{fig:filtering_all}d) illustrates. Across various kinematical conditions, the heavy residue has a small outgoing angle, often resulting in it being identified as part of the beam cluster. An initial step breaks the beam cluster, considering all points with X greater than the preliminary RP just found as heavy recoil (\textbf{1} in \autoref{fig:filtering_all}d).

Next, all points within a sphere-like mask around the RP (\textbf{2}) are deleted from the tracks, to avoid biases owing to the different transformations the clusters have undergone, which might have left tiny parts of the residues attached to the beam track. A further cleansing consists of deleting all points outside a cylinder defined along the cluster linear fit (\textbf{3}). This ensures no noise contributes to it. Finally, the begin and end caps of the tracks are also deleted within a small distance to ensure the track is symmetric (\textbf{4}), which is particularly important for tracks going towards the pad plane boundaries \cite{roger2011}.

The fine treatment concludes by recalculating the reaction point. Since all recoiling tracks (including the heavy one) now contribute to the calculation, only the track maintaining the largest angle with respect to the beam direction is considered, as the uncertainty would increase for low-leaning angles. 

\section{Track reconstruction}\label{sec:trackrec}
The filtered tracks are employed in the \textit{merger} algorithm to produce a minimal physics output. The merger is initiated by applying a set of gates to determine whether the event is of interest. These conditions are read from the configuration file and allow flexibility among different experiments. They usually consist of trigger tagging checks (where only events with a given trigger value are analysed), silicon multiplicities (typically allowing only one hit per silicon layer) and, lastly, track multiplicities. Since we want to study binary reactions, only track multiplicity of three, where the beam and the two ejectiles are measured, is allowed. 

The next action involves establishing which residue impacted the silicon array, which typically corresponds to the lighter particle of the two residues. Its information will be later used in the missing-mass method to reconstruct the excitation energy and the centre-of-mass angle. The light track is identified by computing the two angles with respect to the beam direction. Depending on the particular reaction and the kinematical regime, one relation holds for all events: either $\theta_{\text{light}} < \theta_{\text{heavy}}$ or vice-versa. As an example, taking pick-up transfer reactions, such as $\iso{20}{O}(\iso{}{d}, \iso{3}{He})$, the light particle would maintain the largest angle of the two.
\begin{figure}[!htb]
    \centering
    \includegraphics[width=1\linewidth]{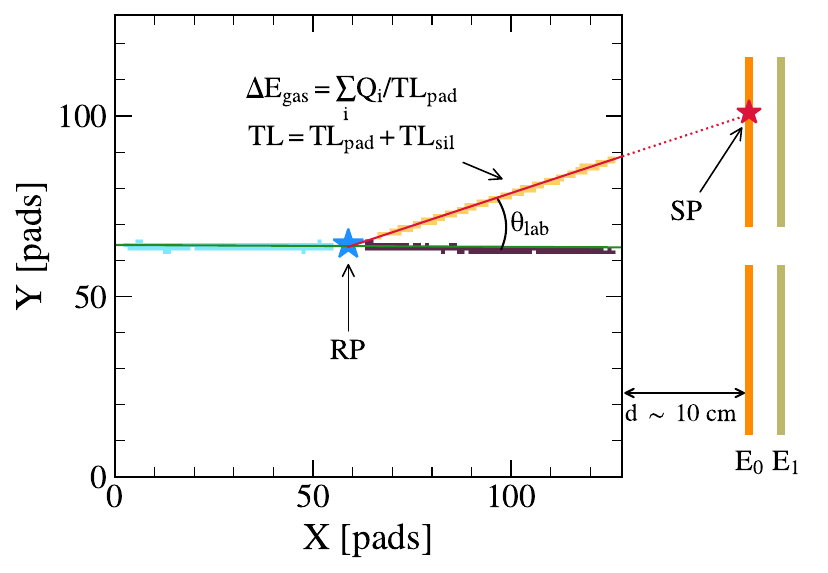}
    \caption{Selection of physical variables computed in the \textit{merger} detector.}
    \label{fig:merger}
\end{figure}

By providing a detailed map of the different silicon detectors (distances to the origin of coordinates and exact positions for each pad), the silicon impact point (SP) of the track is computed. With it, the track length is defined as the distance between the SP and the RP as illustrated in Fig. \ref{fig:merger}. Another variable, essential to perform the Particle Identification (PID), is constructed at this moment: the gas energy loss $\Delta\text{E}_{\text{gas}}$, which is the ratio between the summed track charge $\sum_i Q_i$ and the track length contained within the active region, $\text{TL}_{\text{pad}}$. 

At this point, the drift velocity enters the calculation to convert the $Z$ dimension from time buckets to \unit{\mm} units. If no drift velocity is provided in the configuration file, the \textit{merger} halts the calculations at this step, as no physical angles can be computed (since they would mix different scales along the axes). However, the output is still usable, mainly to determine a first guess for the drift velocity: by gating on one silicon pad and plotting its SP's $Z$ component, the user can determine this parameter by knowing the aforementioned $f_{\text{write}}$ and applying a scaling factor. 

Once a drift velocity is provided, either from experimental measurements or simulations, the \textit{merger} converts all the outputs to \unit{\mm} units and employs them to calculate the physical angles for all the particles, including the beam. Finally, leveraging the fact that the user is asked to provide mappings for the silicon detectors, cuts on the \textit{physical} SP can be applied to mask reactions which have been poorly reconstructed or are not correlated with the silicon trigger, thereby reducing the background.

Lastly, the post-analysis step, where the physics of interest is extracted from the minimal output produced by \textsc{actroot}, can be executed in separate \textsc{root} macros, allowing flexibility in changing experimental conditions. A typical analysis pipeline implies generating the PID plots and then, once particles have been identified, energy-loss corrections can be applied by considering the full track length $\text{TL}$, with the excitation energy $E_{\text{x}}$ and $\theta_{\text{CM}}$ reconstructed in the last step. An example can be found in the \textit{Examples} folder of \cite{zenodo_actroot}. The next section provides an overview of selected physics results and evaluates the program's performance.

\section{Experimental validation}\label{sec:validation}
\subsection{Time comsumption}
Using experimental data from the \iso{20}{O} experiment, \textsc{actroot} was extensively tested with all the data runs. The next table details the time consumption per step, as discussed in Sec. \ref{sec:implementation}, for one average run with \qty{\sim 37000}{} entries and \qty{460}{MB}-sized. The testing machine comprised a standard setup with an \textit{Intel i7 11th generation} CPU and 16 \unit{\giga\byte} of RAM.
\begin{table}[!htb]
\centering
\begin{threeparttable}
\begin{tabular}{@{}ccccc@{}}
\toprule
Step & Cluster & Calibration & Filter & Merger \\ \midrule
Time [s] & \qty{19}{} & \qty{8}{} & \qty{7}{} & \qty{2}{} \\ \bottomrule
\end{tabular}

\end{threeparttable}
\caption{Time usage for a standard data run, covering all steps in the algorithm.
}
\label{tab:performance}
\end{table}

The bottleneck in the analysis workflow is the \textit{cluster} step due to the high-dimensionality of the voxel cloud. Table \ref{tab:performance} lists the execution times using the previously discussed \textit{Continuity} algorithm; if one uses the RANSAC approach, a \qty{3.5}{} factor increase in this key step was measured. 
This highlights the present algorithm's suitability for providing a fast analysis, a crucial factor when executing the online analysis of an experiment. To further improve the processing of large datasets (usually split into \textit{runs}), a multi-threading pool has been provided within \textsc{actroot}, allowing the user to analyse multiple runs in parallel, thereby reducing the execution time by $N$ processing threads.

\subsection{Physics output}
\begin{figure*}[!htb]
    \centering
    \includegraphics[width=1\linewidth]{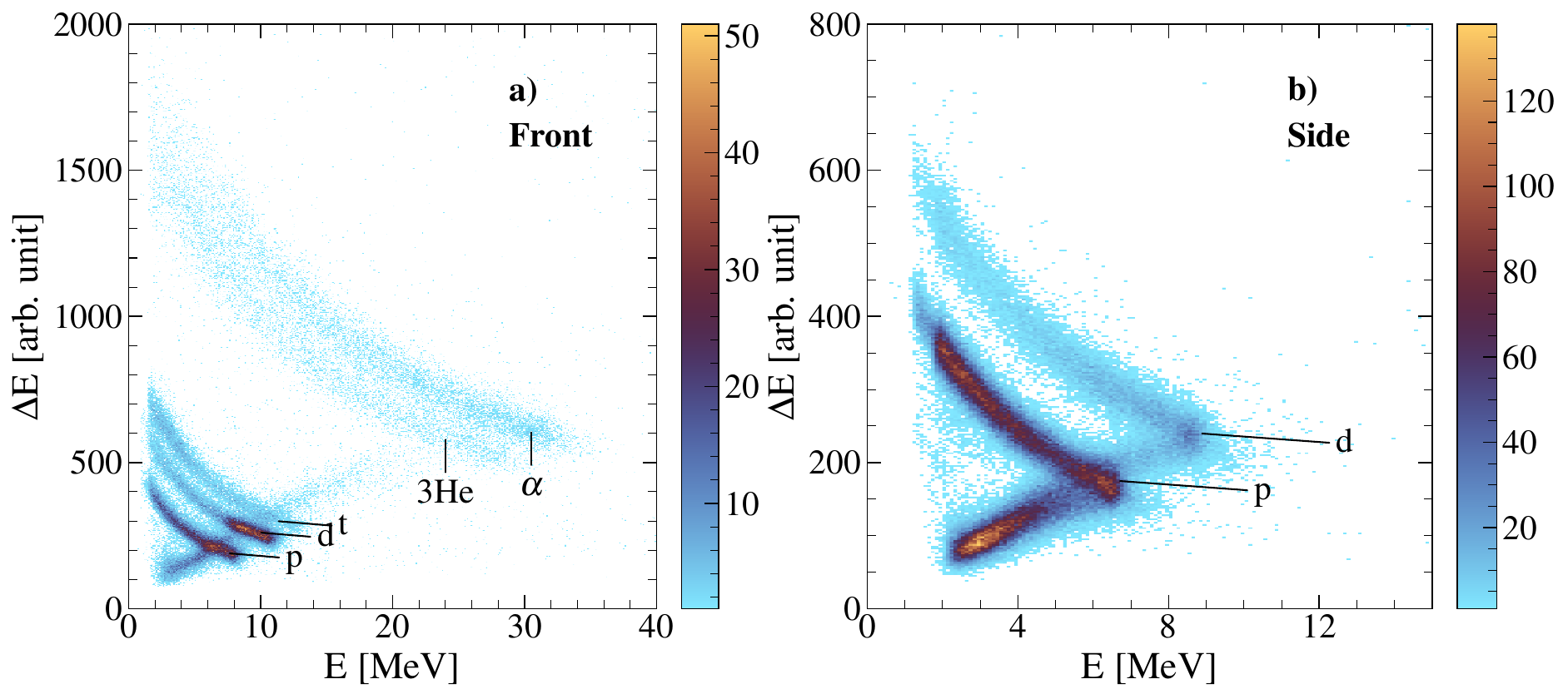}  
    \caption{PID achieved by plotting the measured silicon energy (E) against the average energy-loss in the gas ($\Delta$E) for the two kinematical regimes: \textbf{a)} for the front silicons and \textbf{b)} for the left-side layer.}
    \label{fig:pid}
\end{figure*}

As shown in Fig. \ref{fig:pid}, a clean particle identification matrix is obtained, enabling the separation of hydrogen and helium-like isotopes. This constitutes a strong benchmark of the methods presented in this article. In this figure, the left panel presents the PID for the multiple transfer channels, such as $(\iso{}{d},\iso{}{t})$ or $(\iso{}{d},\iso{3}{He})$, where punch-through events to the second front layer have been masked. On the other hand, the right panel focuses on the left side layer, with the (in)elastic $(\iso{}{p}, \iso{}{p})$ and $(\iso{}{d},\iso{}{d})$ channels reaching it.

\subsection{Reconstruction efficiency}
To assess the algorithm's performance, the reconstruction efficiency for binary reactions must be provided. There is no straightforward way of extracting it without relying on fully-fledged simulations that produce TPC tracks, which are extremely difficult to implement and so far highly dependent on the experimental configuration (gas properties, amplification settings, among others). Consequently, another approach employing the experimental data is pursued: by visually inspecting a representative sample of the events and checking whether the binary events have been correctly identified, the efficiency can be inferred. Due to the low prevalence of real binary reactions in the dataset, achieving low uncertainties is a time-demanding task. For this publication, statistical uncertainties have been kept below \qty{8}{\percent}. Results are summed up in Table \ref{tab:eff}. 
\begin{table}[!htb]
\centering
\begin{tabular}{@{}cSS@{}}
\toprule
            & {\textbf{Side}} & {\textbf{Front}} \\ \midrule
\# inspected   &  1396    &  1909     \\
\% binaries &   43.77   &   8.70    \\
\% reconstructed      &   \num{90.8(2.2:2.6)}   &  \num{96.4(2.3:4.1)}     \\ \bottomrule
\end{tabular}
\label{tab:eff}
\caption{Human-labelled reconstructed efficiency for the two sets of data. For the tracks going to the side layer, an effective reconstruction of 100\% is achieved when low-energy events are required (particles that stop in the Si detectors). Uncertainties (at $2\sigma$ level) were calculated according to the Clopper-Pearson method.}
\end{table}

The overall efficiency is greater than \qty{90}{\percent} in both regions, which demonstrates the algorithm's performance over a broad event topology. Even so, intending to improve the efficiency in the future, the following lines detail the major sources of faults in the reconstruction.

On the one hand, side events should present no difficulties to the algorithm, given their long track lengths and large angles. However, another effect comes into play: high-energy events with small energy deposition in the gas, barely reaching the pad readout threshold, hence resulting in fragmented tracks. The Continuity clustering algorithm is unable to recover them owing to the fragmentation, which causes them not to surpass the minimum voxel requirement. This has indeed been confirmed in an event-specific analysis, proving that those missed events correspond to high-energy light particles punching through the Si detectors. With this in mind, if the analysis is carried out with the low-energy branch ($E$ in the \qtyrange{2}{8}{\MeV} range), the efficiency is expected to be \qty{100}{\percent}. Alternatively, RANSAC might be considered as a more reliable option to recover those high-energy tracks, as preliminary results indicate.

On the other hand, front events are more challenging due to the diversity of event topologies, particularly when the light residue emerges at a low angle with respect to the beam direction, and the reaction takes place at the end of the TPC. To ensure more stable parameters in the algorithm, reactions occurring either at the very beginning of the chamber or towards its end have been excluded from the analysis, effectively excluding vertices outside $\left[40, 200\right] \unit{\mm}$ along the beam direction. Despite this filtering, the \qty{4}{\percent} of missed events was found to arise from  $\delta e^{-}$ contamination, whose large clusters exceed the voxel threshold and artificially create false multiplicities, as \autoref{fig:example_events} illustrates. This level of accuracy is still acceptable, as there is no practical way of defining a universal parameter set suitable for all event types. In any case, the efficiency obtained in this experiment corresponds to a lower limit, as a high-gain was set in the pad plane, thereby enhancing $\delta e^{-}$ observation.

\section{Conclusions}
A complete framework for ACTAR TPC analysis has been developed, and its performance has been evaluated. After motivating the importance of active targets in widening the present knowledge on exotic nuclear structure, the analysis challenges posed by such exotic experiments were presented, and feasible solutions were provided. The \textsc{actroot} framework was proposed as the solution for this device, aiming to be user-friendly and efficient.

In the subsequent sections, the code structure was introduced, and careful attention was paid to describing its multiple components and processing steps. In particular, a fast Continuity clustering algorithm, specially tailored for TPC voxel clouds, was presented. To address the specific characteristics of these devices, such as high background levels and significant track fragmentation, several transformations operating on the resulting clusters were implemented, enabling the extraction of a minimal physics output with reduced uncertainties.

We concluded by showing experimental results for the first transfer experiment conducted with ACTAR TPC employing a $\iso{20}{O}$ beam on a deuterium and hydrogen target at GANIL. Both the PID and reconstruction efficiency confirm \textsc{actroot} robustness for analysing transfer experiments with active targets. Furthermore, the execution time was shown to be considerably lower compared to the data taking itself, enabling online monitoring during experimental campaigns.

\section*{Acknowledgements}
M.L.G. and J.L.F. wish to acknowledge financial support from the Xunta de Galicia (Spain) grant numbers ED481A-2022/419 and ED481A-2020/069. The Spanish MINECO supported this work through the project PID2021-128487NB-I00. This work was
also partially funded by the Xunta de Galicia under projects 2021-PG045
and 2024-PG040, and the Maria de Maeztu Unit of Excellence CEX2023-
001318-M.

\appendix
\section{Data availability}
Experimental data can be found in Reference \cite{e796_19_2022}.



\bibliography{biblio.bib}





\end{document}